\documentclass[a4paper]{article}
\usepackage[utf8]{inputenc}
\usepackage[T1]{fontenc}
\usepackage{amsmath,amssymb,array}
\usepackage{booktabs}
\usepackage{float}
\usepackage{hyperref}
\usepackage{graphicx}
\usepackage{natbib}
\usepackage{authblk}
\usepackage{fancyvrb}
\usepackage{geometry}
\newcommand{\CRANpkg}[1]{\href{http://CRAN.R-project.org /package=#1}{\textsf{#1}}}%

\title{Use of Simulation Models for the Development of a Statistical Production Framework for Mobile Network Data with the simutils Package\footnote{Presented at the Conference Use of R in Official Statistics 2021,  24-26 November 2021, Bucharest (Romania).}}

\author[1]{B. Oancea}
\author[2,3]{D. Salgado}
\author[2]{S. Barrag\'{a}n}
\author[4]{M. Necula}
\affil[1]{Dept. Applied Economics and Quantitative Analysis, University of Bucharest \protect\\
	90, Panduri Street \protect\\ Sector 5, 050663 Bucharest, Romania}
\affil[2]{Dept. Methodology and Development of Statistical Production, Statistics Spain (INE)\protect\\
	Avda. Manoteras, 52 \protect\\ 28050 Madrid, Spain}
\affil[3]{Dept. Statistics and Operations Research, Complutense University of Madrid\protect\\
	Pza. Ciencias, 3 - Ciudad Universitaria \protect\\ 28040 Madrid, Spain}
\affil[4]{Dept. Innovative Tools in Official Statistics, National Institute of Statistics (INS)\protect\\
	16 Libertatii Bvd.\protect\\ Sector 5, 050706 Bucharest, Romania}

\begin{document}

\maketitle

\begin{abstract}
	We propose to use agent-based simulation models for the development of statistical methods in Official Statistics, especially in relation with the new digital data sources. We present a mobile network data simulator which is managed through the \textsf{simutils} R package which provides geospatial representations of the simulated data. While the synthetic data are produced by an external tool, our \textsf{simutils} package allows an R user to parameterize and run this external simulation tool, to build geospatial data structures from the simulation output or to compute several aggregates. The geospatial data structures were designed with the purpose of using them in a visualization package too.
	Useful simulation models require the incorporation of real metadata from mobile telecommunication networks driving us to the inclusion of functionalities allowing the user to specify and validate them. All metadata are specified using XML file whose structure are defined in corresponding XSD files.  
	Our R package includes example data sets and we show here how validate the metadata, how to run a simulation and how build the geospatial data structures and how to compute different aggregates.
\end{abstract}

\section{Introduction}

\subsection{Background}
During the last fifteen years the production of official statistics in national and international statistical offices is facing a double challenge, namely the modernization and industrialization of the production process and the incorporation of new digital data sources. The first challenge was clearly identified in the first decade of the present century \citep{HLG11a} and gave rise to the advent of international production models such as the GSBPM \citep{GSBPM51}, the GSIM \citep{GSIM12}, and the GAMSO \citep{GAMSO12}, to name a few. These standards are being gradually adopted by most statistical offices thus conforming a common working space in the international community \citep{HLG21}. The second challenge was already identified eight years ago \citep[see e.g.][]{SchMem13a} and is taking efforts in the form of complementary institutional initiatives \citep[see e.g.][]{ESSVIP, EurBD21a, UNBD21a}. This challenge \textquotedblleft will require amendments to the statistical business architecture, processes, production models, IT infrastructures, methodological and quality frameworks, and the corresponding governance structures [\dots]\textquotedblright\ \citep{BucMem18a}, i.e.\ basically to modify all aspects of the entire production framework.\\

Multiple analyses have been conducted regarding the use of new digital data for the production of official statistics \citep[see][and multiple references therein]{Kit15a, Han18a, SalOan20a}. Most of them are still far from being solved. Apart from the identification of concrete issues such as the data access, the new methodological and quality assurance frameworks, and the technological environment, more entagled implications arise because of their interaction. More specifically, these new data sources require new statistical methods in the production of official statistics partially abandoning and hopefully integrating the use of sampling designs \citep{SarSweWre92a} with statistical models \citep{ValDorRoy00a}, especially in the fields of statistical learning \citep{HasTbiFri09a,Mur13a}, Bayesian modelling \citep{GelCarSteDunVehRub13a} or geostatistics \citep{SchGot05a}. This new official statistical methodology is developing too slow, among other things, because the access to this sort of data is proving to be extremely hard and new legislation will need to be put into force \citep{BalRicWir21a}. Thus, strategies to boost the development of new statistical production frameworks in this context are needed. As we shall explain in the following sections, the use of simulation models and their analysis with R brings an excellent scenario to face this situation.

\subsection{The contribution of this paper}
\begin{figure}[H]
	\centering
	\includegraphics[width = \textwidth]{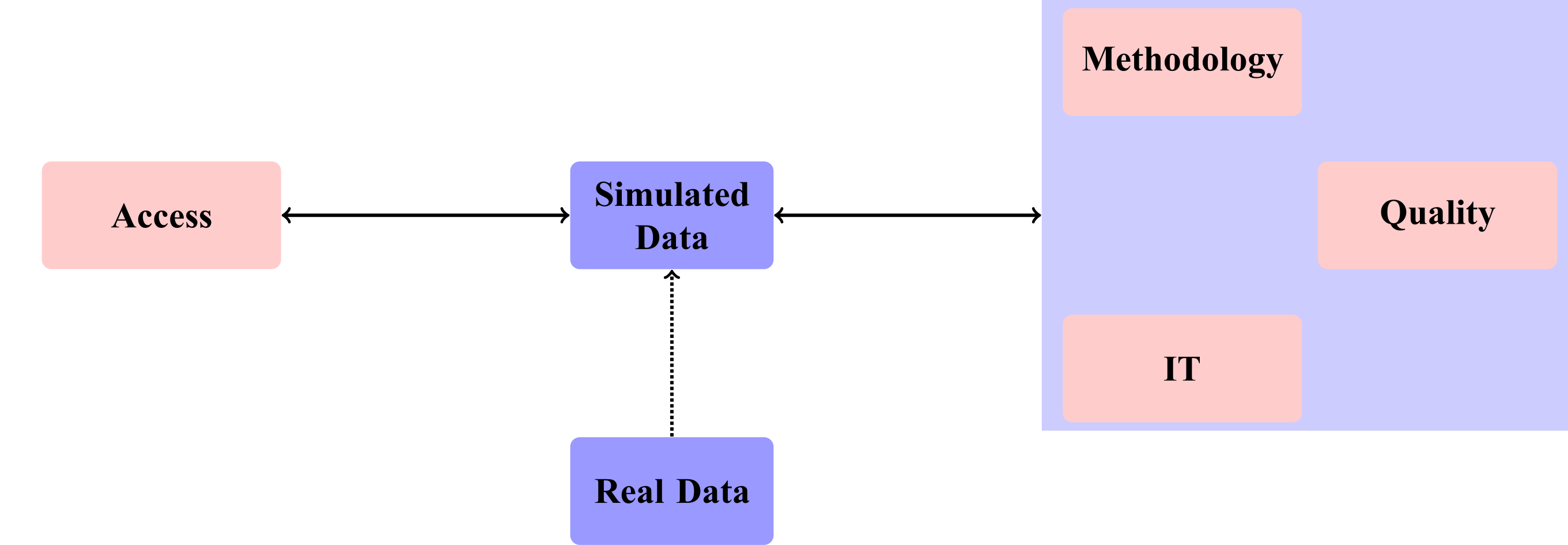}
	\caption{Schematic representation of the strategy to develop and evolve a statistical production framework.}
	\label{figure:strategy}
\end{figure}

A basic tenet of our present contribution is that simulation models producing synthetic data incorporating real metadata must become a fundamental tool to develop these new methodologies and to assess their potentiality. Simulated data must help develop the methodological and quality frameworks as well as assess the technological environment to device a new statistical production chain. This should be combined with the use of real data in such a way that synthetic data are to be finally substituted by them to constitute the ever evolving production cycle. This idea is illustrated in figure~\ref{figure:strategy}. Even when having access to real data, simulated data provides an analytical and testing playground of different aspects of any kind of statistical model used to produce the official statistics.\\

As we shall argue below, the use of R stands as an excellent statistical software development tool to implement this strategy. Following \citet{Cha09c}, this versatility arises mainly because R provides outstanding characteristics for the development and evolution of statistical production frameworks: (i) it is an \emph{interface} to computational procedures of many kinds, (ii) it is \emph{interactive}, (iii) it incorporates \emph{functional} programming facilities, (iv) it incorporates \emph{object-oriented} programming facilities, (v) it is \emph{modular}, and (vi) it is \emph{collaborative}. The need to interact with other computing languages and platforms is beyond doubt a need for any modern computing system. The interactivity allows the user to check in a fast and seamless way the results of intermediate computations. Functional and object-oriented paradigms are essential ingredients in a computing system to process statistical data and to produce reliable statistics. Modularity is the key feature for a scalable system to cope with complexity. The collaboration of the international community in an open environment is fundamental to reach widely adopted production standards. There already exists an important initiative to share open source statistical software for the production of official statistics  \citep{BosLooKow20a}.\\

In this contribution we focus on the use of mobile network data as a novel promising data source for the production of official statistics. This data source is indeed facing a shortage of data access for Official Statistics due to intricate reasons, thus the use of synthetic data will allow us to move forward in the development of statistical methodology.\\

The work is organised as follows. Firstly, we describe from a statistical point of view the complex data ecosystem of mobile telecommunication networks for the production of statistics. Next, we provide generic information about the use of simulation models, especially in connection with mobile network data. Then, we describe the general lines of the statistical software development with the creation of several R packages to implement a modular end-to-end process to produce population counts and origin-destination matrices from this data source. We focus on the package \textsf{simutils}, which provides a set of functionalities to manage simulated mobile network data. We conclude with a brief summary and discussion.

\section{Ecosystem of mobile network data for statistical production}
\label{sec:MND}
Mobile telecommunication networks are extraordinarily complex systems providing communication between moving agents through electromagnetic interactions \citep[see e.g.][]{MiaZanSunSli16a}. These networks daily produce a huge amount of digital data to make this communication feasible and to allow engineers monitor the situation to provide an increasingly higher-quality communication service. As it is widely known and perceived, this kind of communication involves personal data to some extent (in order to establish the desired contact) and, thus, its use for producing official statistics raises more often than not some concern. In this line of thought, it is mandatory to explain as clear as possible what data are to be used to produce what statistics.\\

Firstly, data to be used are generated in the network systems, never ever in the mobile devices. That is to say, the goal is to reuse the digital trace left behind in the mobile telecommunication networks by the mobile devices, i.e.\ by individuals of a given target population of analysis. Hence, the more appropriate expression \emph{mobile network data} instead of \emph{mobile phone data}.\\

Secondly, what should be understood by \textquotedblleft digital trace\textquotedblright? Nowadays two main usages of mobile network data are considered, namely geolocation of mobile devices \citep[see][for a pioneering work and an extensive list of references, respectively]{AhaAasSilTir07a, SalSanOanBarNec21a} and Internet usage \citep{UcaGraFioSmoMor21a} both for sociodemographic analyses. In the former case, telecommunication variables driving us to an estimation of the position of devices through some statistical model are needed \citep{MunBouVarEnrCal09}. In the latter case, network variables showing mobile digital usage are needed \citep{UcaGraFioSmoMor21a}. As we shall argue, analyses with synthetic data may allow us to identify those variables needed to reach the accuracy determined by official statistics quality standards.\\

Thirdly, should we use individual or aggregated data? For official statistics purposes, this is a misleading question. Aggregated mobile network data, if any, arises from a computational procedure upon individual mobile network data. This computational procedure is of statistical nature, especially regarding the geolocation of mobile devices and more intensively when referred to administrative territorial units (city districts, municipalities, \dots). This procedure is executed by someone in a given computing platform following a previously designed algorithm. Thus, the starting point is always individual mobile network data since the network provides service at an individual level. The question should be: are official statistics to be based on some form of raw digital data or on some form of preprocessed statistics (i.e.\ of mobile network statistics, we should say)? This question is far from having a definitive answer and touches the role of Official Statistics in society. In this work we concentrate on the proposal of executing in-situ production processes hopefully designed jointly by statistical officers and telecommunication experts \citep{SalSanOanBarNec21a}.\\

In this line of reasoning we shall work with network event data, i.e.\ data generated due to the interaction between a mobile device and the network. We shall denote this data by $\mathbf{E}_{kt}$, where $k$ denotes the mobile device ID and $t$ denotes the timestamp generated by the interaction (with a precision of seconds). Complementarily, we shall also make use of telecommunication variables about the configuration of the network such as the position of each antenna, their emission power, their path loss exponent, \dots \citep{SalSanOanBarNec21a}. We shall denote this data by $\mathbf{N}_{a}$, where $a$ stands for the antenna\footnote{We assume for simplicity's sake that the configuration of the network does not change during the period of analysis.}. We used the simplified term \textit{antenna} for what mobile network engineers cal a Base Transceiver Station. These data allow us to compute an estimation of the signal strength from each antenna or a similar measure in each tile of a territorial grid of analysis through a so-called radio propagation model \citep{MiaZanSunSli16a}. Notice that the information contained in $\mathbf{E}_{kt}$ will be sensitive for users whereas the information contained in $\mathbf{N}_{a}$ will be sensitive for mobile network operators (hence the intricate complexity of access to this data source).\\

\begin{figure}[H]
	\centering
	\includegraphics[width = \textwidth]{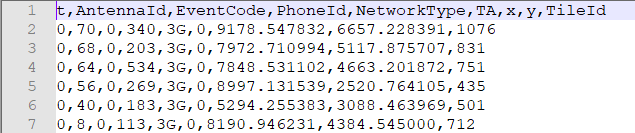}
	\caption{Example of simulated network event data.}
	\label{figure:eventDataExample}
\end{figure}

Network event data can comprise multiple telecommunication variables, some of them with no utility for our statistical purposes and in constant evolution with the technology. For statistical purposes regarding geolocation, we shall simulate datasets collecting the following information from each network event: (i) the timestamp of the network event, (ii) a constant identification variable for the device during the time period of analysis, (iii) an identification variable for the radio cell where the event is produced, (iv) a code for the type of event, (v) the network technology type (3G, 4G, etc.) and (vi) the \emph{Timing Advance} of the signal (basically a code denoting the time for the signal to reach the antenna from a mobile device). We complement these variables with the simulated real position of the device where the network event is produced (coordinates and grid tile identification). This synthetic ground truth will allow us to assess the performance of any statistical model of geolocation and of estimation of population counts and mobility patterns (see figure \ref{figure:eventDataExample}).

\begin{figure}[H]
	\centering
	\includegraphics[width = \textwidth]{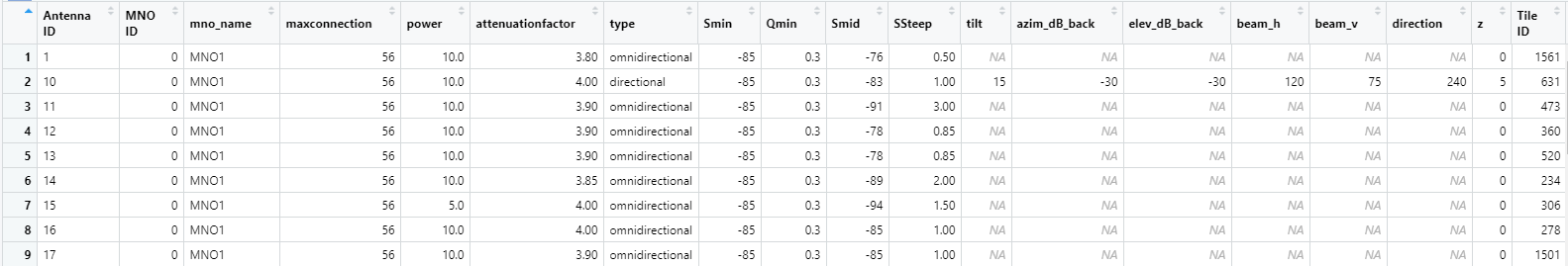}
	\caption{Example of simulated network parameter configuration.}
	\label{figure:networkDataExample}
\end{figure}

Regarding the network configuration parameter datasets, we collect variables regarding (i) antenna (radio cell) identification, (ii) the mobile network operator identification, (iii) the mobile network operator name, (iv) the maximum number of connections per radio cell, (v) the emission power (in W), (vi) the path loss exponent, (vii) the type of radio cell (omnidirectional, 120$^{\circ}$-directional), (viii) the minimum signal strength to establish a connection (in dBm), (ix) the minimum signal dominance (as defined in \citet{TenGoo21a}) to establish a connection, (x) the midpoint parameter for the computation of the signal dominance (in dBm), (xi) the steepness parameter for the computation of the signal dominance, (xii-xviii) diverse orientation parameters for the antenna (tilt, elevation, etc.), and (xix) the grid tile identification of the antenna position (see figure \ref{figure:networkDataExample}).\\

Both network event and network parameter variables are generated through a network event data simulator developed in a modular form for this purpose, as explained below.\\ 

From the perspective of the statistical analysis to be conducted, this wealth of complex and technological data $\mathbf{E}_{kt}$, $\mathbf{N}_{a}$ should not overwhelm the statistician since the ultimate goal is to build a statistical model to compute the location probability $\mathbb{P}\left(\textrm{tile}|\mathbf{E}_{kt}, \mathbf{N}_{a}\right)$ for every single mobile device $k$ at each time instant $t$ of analysis. These location probabilities allow us to detach the analysis from the extremely technological substratum so that deduplication of devices carried by the same individual, identification of target individuals (tourists, commuters, etc.), aggregation to produce probability distributions of population counts, and inference to the whole target population can be undertaken on a statistical basis \citep{TenGooSha20a,TenGoo21a,SalSanOanBarNec21a, OguBenOmeAlo21a}. More sophisticated models involving another telco variables can also be possibly devised, hence the hopeful collaboration between statistical offices and mobile network operators. Synthetic data and comparison with the simulated ground truth will allow the statistician to assess the model performance and the quality indicators (accuracy, bias, etc.).

\section{Data simulation for the development of statistical production frameworks}
\label{sec:Simulation}
A scientific model can be defined as a \textquotedblleft simplified representation of a system under study, which can be used to explore, to understand better or to predict the behaviour of the system it represents\textquotedblright\ \citep{OSul13}. A key word in this definition is \textquotedblleft simplified\textquotedblright, which is a conceptual need arising from the scientific exercise itself of understanding and describing reality. This is also present in the use of any type of data, either simulated or real. As a matter of fact, despite the huge deluge of data inviting to some kind of objective apprehension of reality thereof, this is not so:  some sort of conceptual, mathematical, phsyical, or scientific modelling exercise must be undertaken on these data, all during generation, collection, processing, and inference. Both real and simulated data are \emph{a sociotechnical artefact} \citep{Kit17a}. Hence, the well-known motto \textquotedblleft Essentially, all models are wrong\dots but some are useful\textquotedblright\ by \citet{Box79a} \citep[see also][and articles in the same monographic journal volume number]{WitHeuRom12a}. Notice, however, that we have multiple ways of expressing the goodness of fit of a model (how wrong it is), but we still lack the formal idea of a model utility (how useful it is), despite being a fundamental question \citep[see e.g.][for a concrete discussion related to earthquake risk modelling]{Fie15a}. In our context, simulation models are cleary oriented towards the development of statistical methods driving us to the identification and selection of telecommunication variables, geolocation  methods, and statistical procedures in general producing official statistics of interest for the public good. Official Statistics is not about real data, but about an accurate description of reality for policy-making and decision-taking.\\

\begin{figure}[H]
	\centering
	\includegraphics[width = 0.5\textwidth]{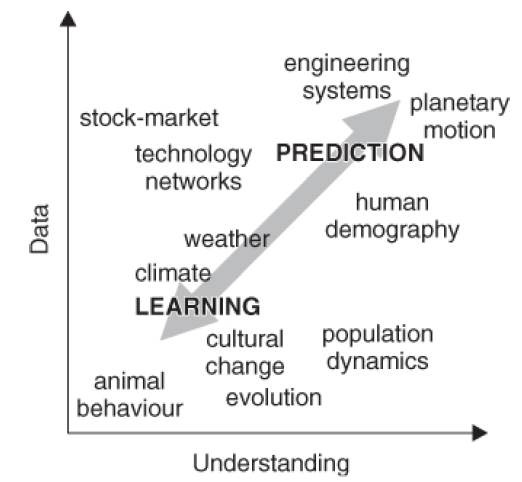}
	\caption{Data versus understanding (from \citet{OSul13})}
	\label{figure:data_understanding}
\end{figure}

This concept of model utility, however it is defined, is related to how scientists use models, which \citet{OSul13} illustrate in terms of two critical aspects, namely data availability and phenomenon understanding (see figure~\ref{figure:data_understanding}). When we lack data and have a poor understanding of a phenomenon, simulation models can be used to gain learning. On the other hand, when both data and understanding abound, we can use models to make accurate predictions. Currently, the production system of official statistics based on mobile network data is clearly in a learning stage, hence the natural use of simulation models.

In this context, agent-based models (ABMs) stand as an excellent tool to explore and test different proposals to create a statistical process based on mobile network data. They are simulation models that describe the behaviour of individual organisms.  In the past, the complexity of these kinds of models caused a lack of use. However, in recent years due to the computational progress they have undergone an expansion. These models are used in a wide range of fields such as Ecology \citep{DeAng18}, Economics \citep{Tes02}, Social Science \citep{Gil05} and much more. The importance of ABMs is obvious since they allow to represent complex systems and analyse them in multiple and different scenarios. Even more in the case of the development of statistical methodology, these models are essential to assess the quality of the process and make comparisons among different methods.\\

Simulated populations have already been traditionally used in Official Statistics (e.g.\ in survey sampling to assess the performance of finite-population design-based estimators). However, ABMs are not an extended tool, among other things due to their complexity, to computational issues, and to interpretation challenges. This is not exclusive in Official Statistics and many other fields also face the same problem. With the aim of facilitating the use and understanding of these models a standard protocol was proposed some years ago \citep{Grimm06}. This protocol basically conveys the key elements of an ABM by making explicit seven items grouped in three blocks: (i) overview (purpose, state variables and scales, process overview and scheduling), (ii) design concepts, (iii) details (initialization, input, submodels). The idea is to present first the general considerations, then the strategy and finally the technical details \citep[see][for details]{Grimm06}. If ABMs are to be used in Official Statistics, this protocol or an adaptation thereof should be agreed and followed.\\ 

There exists a methodology to create simulation models \citep[see e.g.][and multiple references therein]{Law08a} but, in our view, there exists a fundamental aspect in their use for Official Statistics both from a technical and a strategic point of view. Despite being necessarily a simplification of the final production system, simulation models must incorporate \emph{real} metadata into their construction so that simulated agents and their interacting context resemble as much as possible real-life conditions. This provides a minimal guarantee of reproducing some aspects of reality. However, the goal is to develop versatile and robust statistical methods capable of reliably estimating the synthetic ground truth, simulated with multiple parameterizations of the metadata. If a new ground truth is simulated, the statistical model should be able to estimate it. This is also paramount for the assessment of the quality framework for the devised statistical process.\\

This incorporation of metadata, in our view, should be an essential element for the collaboration between statistical offices and mobile network operators, both at the development and the execution stages of the incorporation of this data source into the production of official statistics.

\section{simutils: an R package to manage simulated mobile network data}
\label{sec:simutils}

In this paper we show the importance of trustworthy software in the development of statistical production frameworks. As stated by \citet{Cha08a}, \textquotedblleft users of the analysis have no option but to trust the analysis, and by extension the	software that produced it. Both the data analyst and the software provider therefore have a strong responsibility to produce a result that is trustworthy,	and, if possible, one that can be shown to be trustworthy [\dots]\textquotedblright. Moreover, regarding software prototyping, \citet{Dav95} said that "The best way to assess what users really need is to give them a “working” system [\dots]".\\

Bearing these ideas in mind, we have chosen the R language due to the numerous advantages that make it a trustworthy software, for instance: (i) it is \emph{open source}, (ii) it is \emph{user-friendly} for statisticians and data scientists, (iii) it is \emph{agile enough} to allow a reasonable speed \emph{in implementing new prototypes}, (iv) there is a \emph{wide range of packages and functions} well developed and related to statistical analysis, especially the latest improvements on geocomputation functionalities \citep{LovNowMue22a}, apart from the essential feature of taking into account the functional and object-oriented paradigms. 

\subsection{A modular approach: the software stack}

\begin{figure}[H]
	\centering
	\includegraphics[width = 0.6\textwidth]{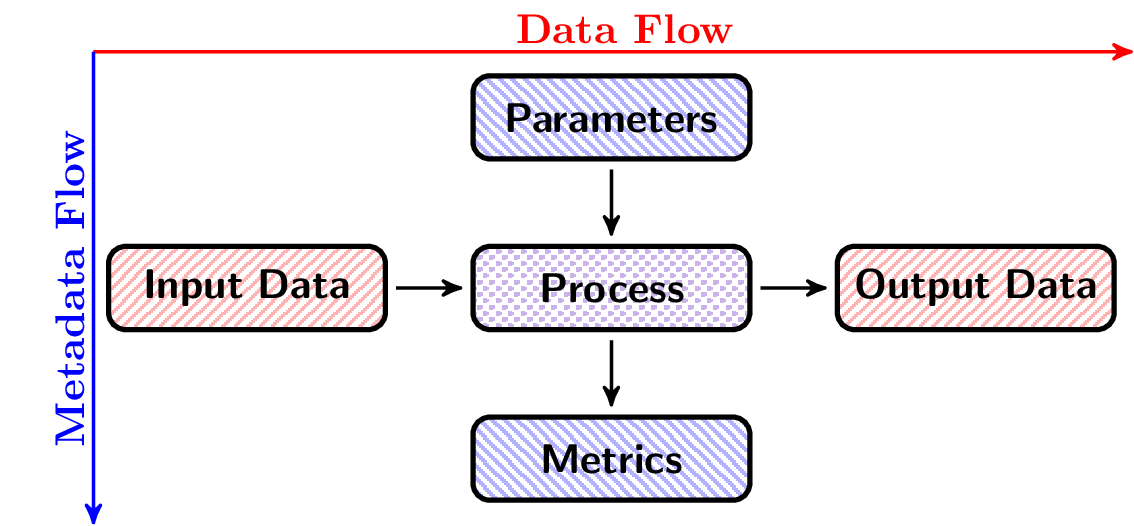}
	\caption{Building block for the statistical process (adapted from \citet{Loo19})}
	\label{figure:dataStep}
\end{figure}

Modularity is a key element in any production process, thus also in statistical production \citep{BalCla00a, Esteban2018}. The idea of modularity is to build the whole process as a set of fully independent modules where any building block follows the generic structure of a statistical production step (see figure~\ref{figure:dataStep}). The modularity allows us to update, modify, and adapt each module according to the statistical domain, statistical needs, and data ecosystem we face in each case. The modularity also allows us to approach the issue about data access with a higher degree of flexibility.\\

The R language is without any doubt a good choice for doing implementation of a modular process as it is presented by \citet{Loo21}.\\

\begin{figure}[h]
	\centering
	\includegraphics[width = \textwidth]{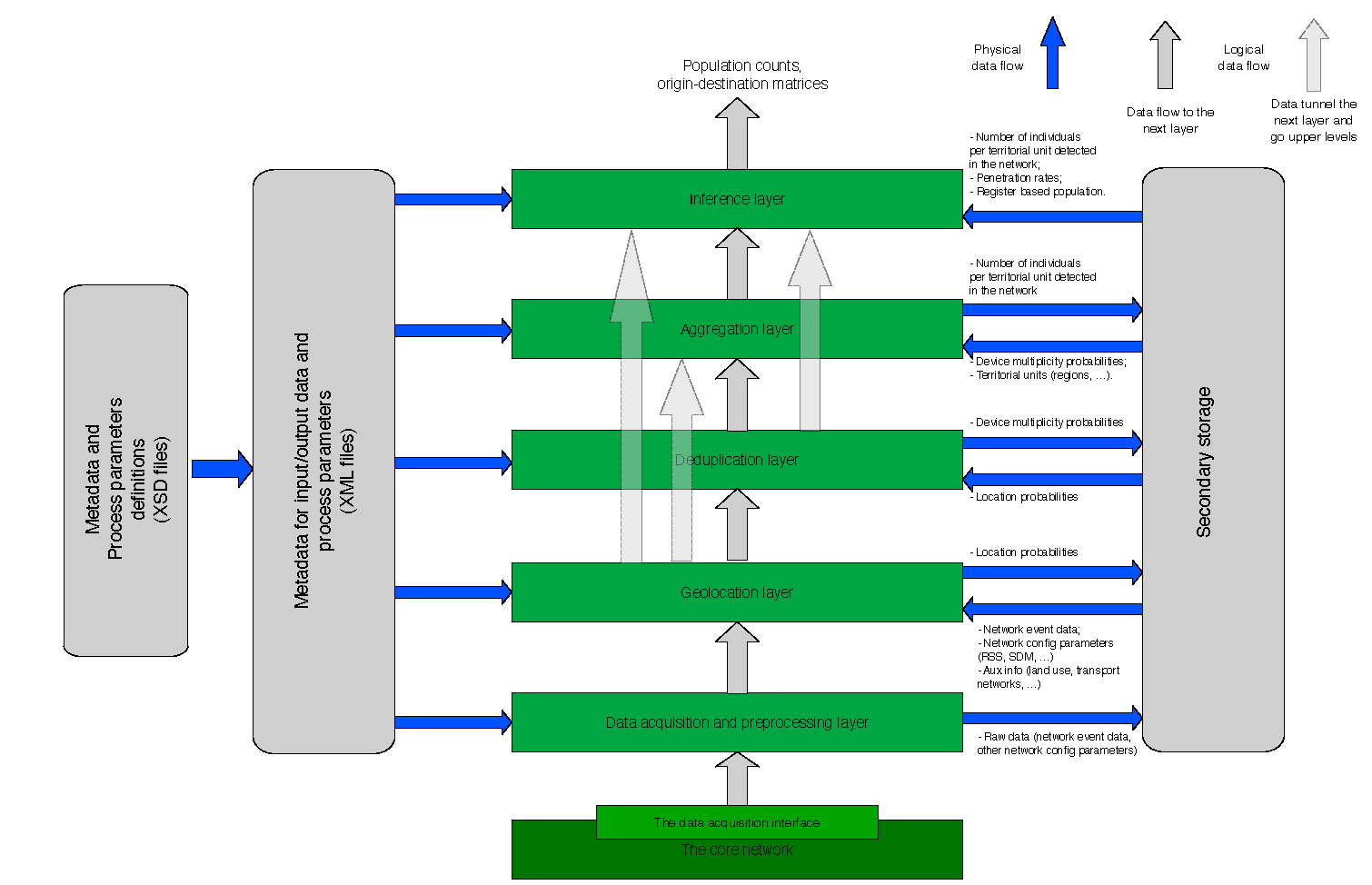}
	\caption{The layered structure of the software implementation (the statistical filtering module is missing because we are estimating the whole population, i.e.\ all individuals are target)}
	\label{figure:layered}
\end{figure}

The big picture of the whole process to analyse mobile network data goes from the raw telecommunication data (real or simulated) to the final estimation of population counts and origin-destination matrices along a fully modular process. The methodology framework of this process has been developed in the European project called ESSnet on Big Data \citep{ESSnetBD21a}. The implementation has been done mostly in R with some preliminary prototypes for each process step. In particular, five big modules were proposed, namely, (i) the geolocation of network events, (ii) the device multiplicity deduplication, (iii) the statistical filtering of target individuals, (iv) the aggregation of mobile devices, and (v) the inference over the target population (see figure~\ref{figure:layered}).\\

In relation to this project the developed software is described by \citet{Oan20}, \citet{oancearsr2021} \citep[see also][]{ESSnetBDMobile21a}. There is a lot of work in progress regarding a general improvement of all the packages presented as prototypes, leaving for future work the issue of scalability for the application in real production conditions.

\subsection{Mobile network data simulator}

\begin{figure}[h]
	\centering
	\includegraphics[width = 1.1\textwidth]{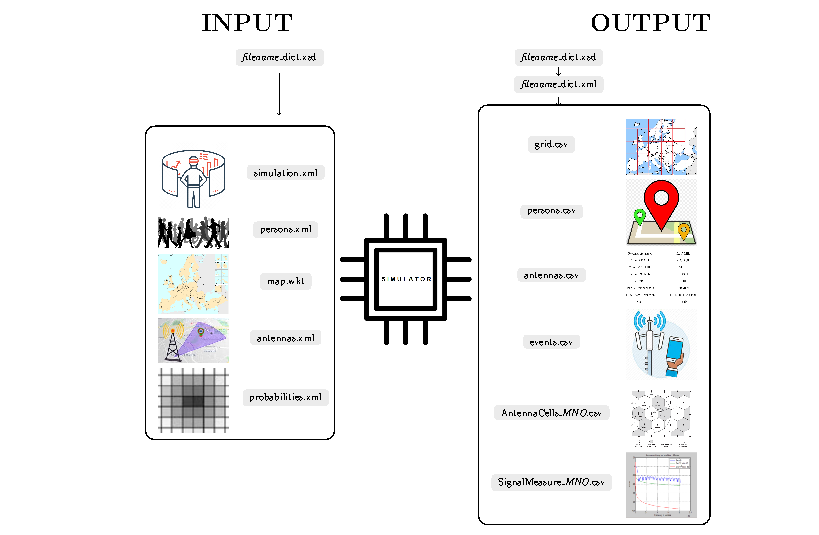}
	\caption{Inputs and Outputs of the simulator}
	\label{figure:MPD_SimDataStructure}
\end{figure}

In the context of the European project ESSnet on Big Data, a mobile network data simulator has been developed in C++ \citep{Oan19, ESSnetBDMobile21a}, whose constant evolution is available as open source software in github \citep{Simulator21a}. This simulator needs a set of input files in XML format and a geographical map or polygon alike in WKT format. These files contain data about (i) the simulation (time span information, mobility patterns, probabilities for individuals of having one or two mobile devices), (ii) persons (how many individuals comprise the population, how fast they walk or move by car, how long they spend at home, at work, etc.), (iii) antennas (their position, emission power, path loss exponent, and other configuration parameters), and (iv) prior specifications for the computation of location probabilities (see figure~\ref{figure:MPD_SimDataStructure}).\\

Once the simulator is executed with all these parameters, the following outputs can be obtained: (i) the grid of analysis in which the geographical territory is divided, (ii) the ground truth about the persons at each time instant (their positions and their devices), (iii) the parameters of each antenna, (iv) the network events arising from the connections of the devices to each antenna at each time instant, (v) the coverage area for each antenna, and (vi) the signal strength (or the signal dominance) of each antenna in the center of each grid tile. In figure~\ref{figure:MPD_SimDataStructure} the whole structure of the simulator's inputs and outputs is represented.\\

The structure of each file is specified in a corresponding dictionary in XML, whose contents, in turn, are specified also in a corresponding XSD file. These file structures allow the user to represent the metadata by using a standard format in XML as well as to conduct their validation with the dictionaries in XSD format. This validation is done using a Java program integrated in the \textsf{simutils} package through the \CRANpkg{rJava} package \citep{pkg:rJava} allowing for compatibility with some features introduced in XML ver. 1.1 (features not yet in R packages dealing with XML processing).

\subsection{The \textsf{simutils} package}

\begin{figure}[H]
	\centering
	\includegraphics[width = \textwidth]{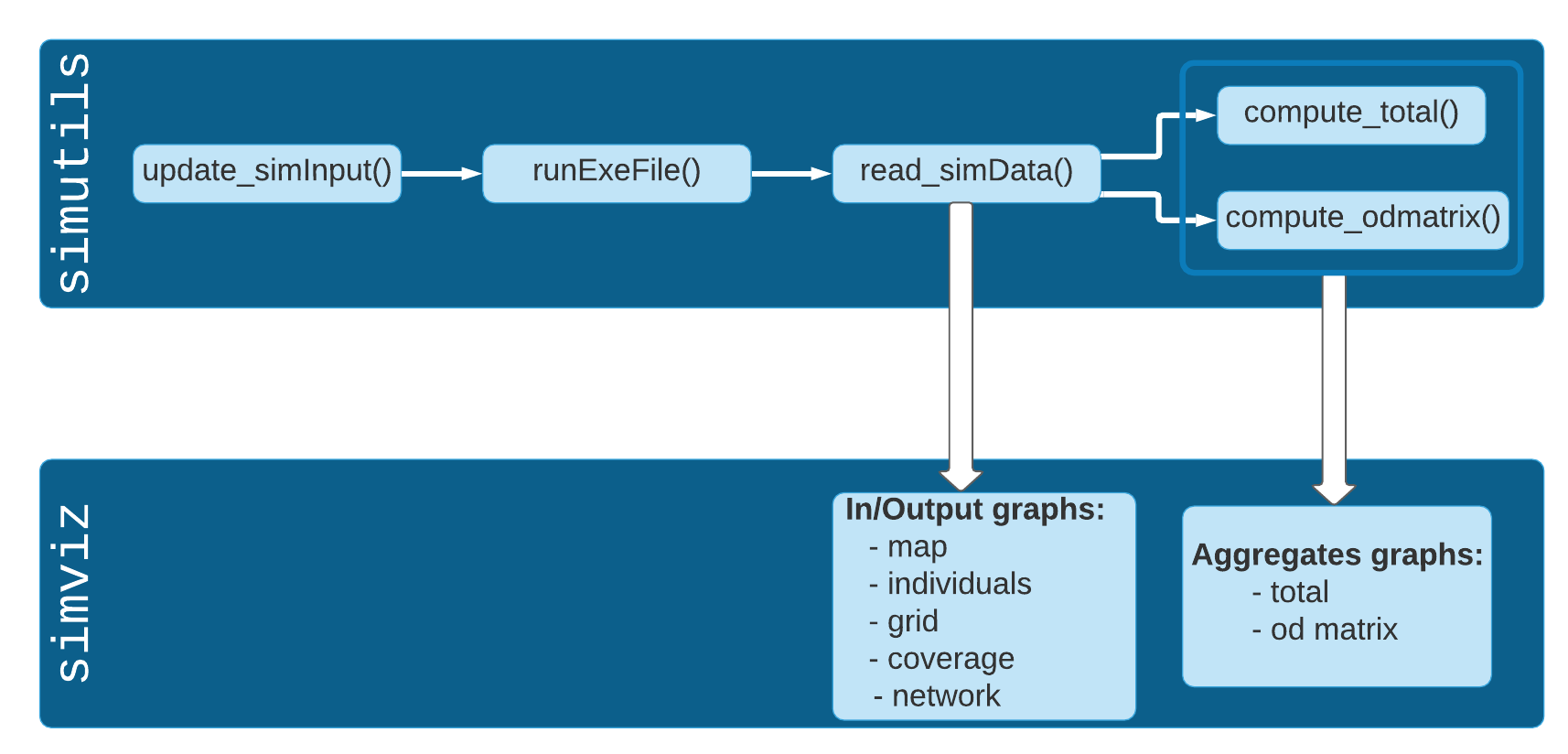}
	\caption{Execution pipeline MND}
	\label{figure:pipeline}
\end{figure}

The execution pipeline to deal with all the information related to the simulator is shown in figure~\ref{figure:pipeline}. Although the simulator has been implemented in C++, all the information related to it can be managed with the R language by using the \textsf{simutils} package, as it is shown in this section. This package is available in github at \url{https://github.com/bogdanoancea/simutils} and is currently under constant evolution. With this tool, it is possible to execute the simulator as well as to manage, specify and change their inputs and outputs. More, the outputs of the  \textsf{simutils} package can be sent to \textsf{simviz}, another R package that we created to build different visualizations of the individual or aggregated data regarding the mobile devices and persons or have a visual representation of different aspects of the mobile network and its characteristics.

Next, we show how to execute step by step the pipeline to obtain simulated datasets as well as some preliminary computations with the ground truth data (total population counts and origin-destination matrices). All files used in the following examples are available together with the \textsf{simutils} package.\\

The first step is to specify and validate the input files for the simulator. This is seamlessly accomplished using the XML/XSD file structures mentioned above:

\begin{Verbatim}
	rootPath <- file.path(system.file(package = "simutils"), 'extdata')  
	
	# Validate the simulation.xml file with its corresponding xsd file
	xml_fn <- file.path(rootPath, "input_files", "simulation.xml")
	xsd_fn <- file.path(rootPath, 
	"metadata/input_files/schema_definition", 
	"simulation_dict.xsd")
	validate_xml(xsd_fn, xml_fn)
	
	# Validate the antennas.xml file with its corresponding xsd file
	xml_fn <- file.path(rootPath, "input_files", "antennas.xml")
	xsd_fn <- file.path(rootPath, 
	"metadata/input_files/schema_definition", 
	"antennas_dict.xsd")
	validate_xml(xsd_fn, xml_fn)
	
\end{Verbatim}

Next, any modification of the simulator input files can be possibly undertaken with the \texttt{update\_simInput()} function:

\begin{Verbatim}
	# rootPath is taken from the preceding code snippet
	xmlSimInput  <- file.path(rootPath, 'extdata/input_files', 'simulation.xml')
	newParam.lst <- list(
	end_time = 11,
	movement_pattern = structure(list(
	manhattan_grid = structure(
	list(x_step = 40, y_step = 40, x_origin = 0, y_origin = 0),
	type = 'home_work_manhattan'))))
	xsdName      <- file.path(
	rootPath, 
	'extdata/metadata/input_files/schema_definition', 
	'simulation_dict.xsd')
	# New simulation input file to be created in working directory
	newSimInputFile <- file.path(getwd(), 'newSimulation.xml') 
	update_SimInput(
	xmlSimInput = xmlSimInput, 
	newParam    = newParam.lst, 
	xsdName     = xsdName,
	newFileName = newSimInputFile)
\end{Verbatim}

The next step is to run the simulation, but firstly we need the proper configuration of the simulator on your computer. The simulator has to be installed separately. It is distributed as source code but we also provide an installation kit.
The source code can be downloaded from  \url{https://github.com/bogdanoancea/simulator/} and the executable file can be built using a C++ compiler. Detailed instructions on how to build the simulator for Windows, MacOs or Linux operating systems are provided on the above mentioned github page. The easiest way to install it under Windows is to download the installation kit from the following link: \url{https://github.com/bogdanoancea/simulator/releases/tag/1.2.0-kit}. Assuming the default installation path, once installed, it is possible to run the simulation from R with the \texttt{runExeFile()} function as follows:

\begin{Verbatim}
	# rootPath is taken from the preceding code snippet
	run_ExeFile(
	path_to_exe       = "C:/Program Files/Simulator",  
	simulator_version = "1.2.0",
	input_folder      = file.path(rootPath, "extdata/input_files"),
	simulationCFGFile = "simulation.xml",
	mapFile           = "map.wkt",
	personsCFGFile    = "persons.xml",
	antennasCFGFile   = "antennas.xml")
\end{Verbatim}

The simulation tool is also distributed as a docker image. If one wants to use the docker image instead of the executable file, the \texttt{runDockerImage()} function should be run:

\begin{Verbatim}
	# rootPath is taken from the preceding code snippet
	input_folder      <- file.path(rootPath, "extdata/input_files")
	output_folder     <- file.path(Sys.getenv('HOME'), 'example_docker')
	run_DockerImage(
	input_folder = input_folder,
	simulationCFGFile = 'simulation.xml',
	mapFile = 'map.wkt',
	personsCFGFile = 'persons.xml',
	antennasCFGFile = 'antennas.xml',
	output_folder = output_folder)
\end{Verbatim}

This function automatically pulls the latest version of the simulator's docker image from the central docker hub.

Once the simulation has been produced, the resulting files can be read with the \texttt{read\_simData()} function as follows. We mention that the \textit{map.xml} file used here augments the initial map of the simulation by adding subdivisions (territorial units) of the territory under consideration. 

\begin{Verbatim}
	filename_map      <- c(
	xml= system.file("extdata/input_files", "map.xml", package = "simutils"),
	xsd= '')
	filename_network  <- c(
	csv= system.file("extdata/output_files/antennas.csv", 
	package = "simutils"),
	xml= system.file("extdata/metadata/output_files/antennas_dict.xml", 
	package = "simutils"))
	filename_signal <- c(
	csv= system.file("extdata/output_files/SignalMeasure_MNO1.csv", 
	package = "simutils"),
	xml= system.file("extdata/metadata/output_files/SignalMeasure_dict.xml", 
	package = "simutils"))
	filename_coverage <- c(
	csv= system.file("extdata/output_files", "AntennaCells_MNO1.csv", 
	package = "simutils"),
	xml= system.file("extdata/metadata/output_files/AntennaCells_dict.xml", 
	package = "simutils"))
	filename_events <- c(
	csv= system.file("extdata/output_files/AntennaInfo_MNO_MNO1.csv", 
	package = "simutils"),
	xml= system.file("extdata/metadata/output_files/events_dict.xml", 
	package = "simutils"))
	filename_grid <- c(
	csv= system.file("extdata/output_files/grid.csv", 
	package = "simutils"),
	xml= system.file("extdata/metadata/output_files/grid_dict.xml", 
	package = "simutils")) 
	filename_individ <- c(
	csv= system.file("extdata/output_files/persons.csv", 
	package = "simutils"),
	xml= system.file("extdata/metadata/output_files/persons_dict.xml", 
	package = "simutils"))   
	
	filenames <- list(
	map                = filename_map,
	network_parameters = filename_network,
	signal             = filename_signal,
	events             = filename_events,
	coverage_cells     = filename_coverage,
	grid               = filename_grid,
	individuals        = filename_individ)
	
	simData <- read_simData(filenames, crs = 2062)
	
\end{Verbatim}

\begin{figure}[H]
	\centering
	\includegraphics[width = \textwidth]{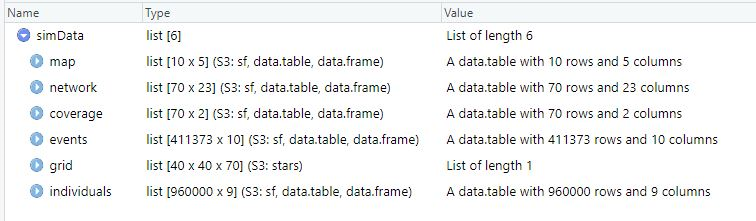}
	\caption{Objects in simData}
	\label{figure:simData}
\end{figure}

The output of this function is the workhorse of the package. It is a list of spatial objects from the \CRANpkg{stars} and \CRANpkg{sf} packages \citep{pkg:stars,pkg:sf} used to represent the territory in the map, the network information, the coverage of each antenna, the network events variables and the signal strength in the grid as well as the ground truth about the individuals. The choice of these packages to represent geospatial objects is complemented with the choice of the \CRANpkg{data.table} package \citep{pkg:data.table} to allow for an efficient, fast and high-volume data management. The structure of this object can be observed in figure~\ref{figure:simData}.

Another functionality of this package is to compute some aggregates of the ground truth (total population counts and origin-destination matrices) that can be computed with the \texttt{compute\_total()} and \texttt{compute\_odmatrix()} functions. The code to obtain each aggregate is included in the following lines:

\begin{itemize}
	\item Counting individuals by territorial unit (\texttt{Subregion\_long} in this example) and time is undertaken as follows:
	
	\begin{Verbatim}
		compute_odmatrix(
		individuals.sf = simData$individuals, 
		what           = 'individuals', 
		by             = c('t', 'Subregion_long'))
	\end{Verbatim}
	
	\item Counting devices by territorial unit (\texttt{Subregion\_long} also in this example) and time:
	\begin{Verbatim}
		compute_odmatrix(
		individuals.sf = simData$individuals, 
		what           = 'devices', 
		by             = c('t', 'Subregion_long'))
	\end{Verbatim}
	
	\item Counting individuals by territorial unit (\texttt{Subregion\_long} also in this example) and time with 0 devices:
	\begin{Verbatim}
		compute_odmatrix(
		individuals.sf = simData$individuals, 
		what           = 'individuals_dev0', 
		by             = c('t', 'Subregion_long'))
	\end{Verbatim}
	
	\item Counting individuals by territorial unit (\texttt{Subregion\_long} also in this example) and time with 1 device:
	\begin{Verbatim}
		compute_odmatrix(
		individuals.sf = simData$individuals, 
		what           = 'individuals_dev1', 
		by             = c('t', 'Subregion_long'))
	\end{Verbatim}
	
	\item Counting individuals by territorial unit (\texttt{Subregion\_long} also in this example) and time with 2 devices:
	\begin{Verbatim}
		compute_total(
		individuals.sf = simData$individuals, 
		what           = 'individuals_dev2', 
		by             = c('t', 'Subregion_long'))
	\end{Verbatim}
	
	\newpage
	\item Counting multiple totals by territorial unit (\texttt{Subregion\_long} also in this example) and time:
	\begin{Verbatim}
		totals <- c('individuals', 'individuals_dev0', 'devices')
		compute_odmatrix(
		individuals.sf = simData$individuals, 
		what           = totals, 
		by             = c('t', 'Subregion_long'))
	\end{Verbatim}
\end{itemize}

\begin{figure}[h]
	\centering
	\includegraphics[width = \textwidth]{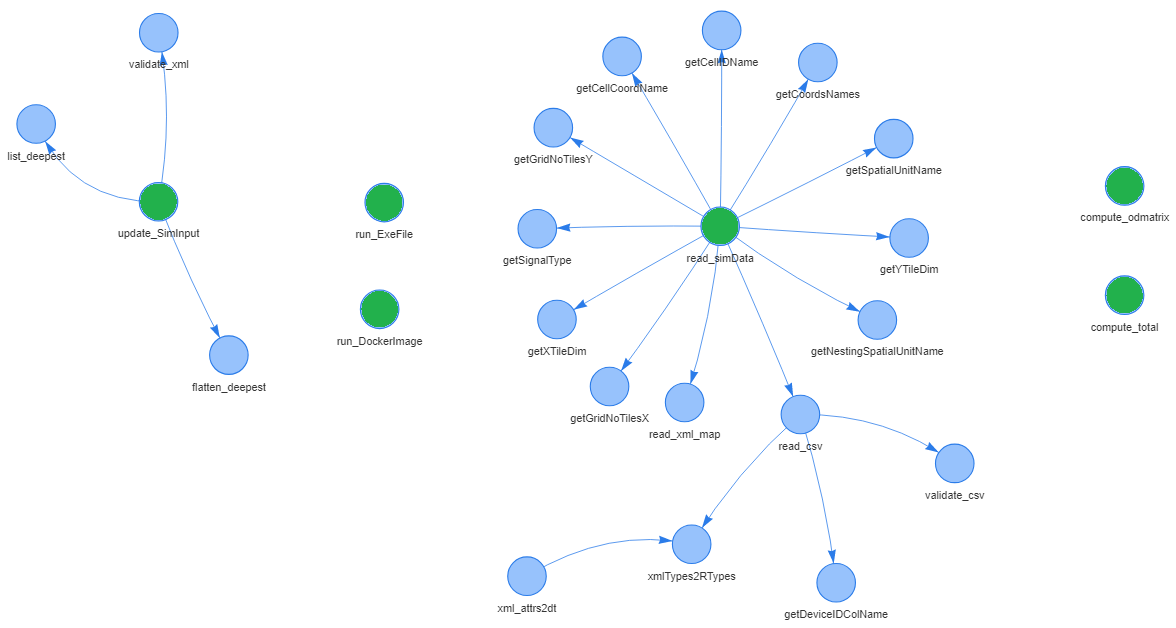}
	\caption{Function dependencies in the \textsf{simUtils} package (drawn with the \textsf{DependenciesGraphs} package \citep{pkg:DependenciesGraphs}).}
	\label{figure:dep_functions}
\end{figure}

The functions described above are the primary functions, but there is a bunch of secondary functions which are needed to deal with the information. In figure~\ref{figure:dep_functions} there is a graph with the dependency relations among all the functions in the \textsf{simutils} package. Primary functions are represented in green.

\begin{figure}[H]
	\centering
	\includegraphics[width = 0.5\textwidth]{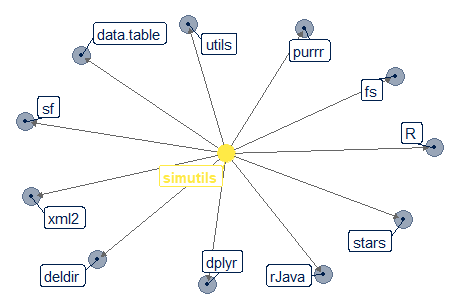}
	\caption{Packages dependencies (drawn with the \textsf{depgraph} package \citep{pkg:depgraph})}
	\label{figure:dep_packages}
\end{figure}

In figure~\ref{figure:dep_packages} there is a graph representing the packages dependencies.

The outputs of \textsf{simutils} can be visualized with the \textsf{simviz} package, which is under early construction. Firstly, we have prioritised the choice and implementation of the representation of these simulated network data (the simulator's inputs and outputs) into R geospatial objects to facilitate next both the statistical analyses and the visualization with standard tools already offered in the R ecosystem. The visualization of these objects will be undertaken in the \textsf{simviz} package making use of the grammar of graphics functionalities providing combined visualizations of the map and their territorial units, the grid, the antenna coverage, the network events and the individuals movements. The computation and statistical analysis (e.g. the geolocation of mobile devices) will be undertaken in other packages \citep[see e.g.][]{pkg:destim}, which we are refactoring to include the aforementioned representation of geospatial objects.

\section{Summary and discussion}
\label{sec:Disc}
The advent of new data sources in Official Statistics is also bringing the need for new statistical methods to infer about target populations with the traditional high-quality standards. This new data sources are mainly digital data with the characteristics of Big Data, 
thus rendering their access, management and processing remarkably complex. In this line, the development of new statistical production 
frameworks becomes thus a challenge, since the use of real data to research stands as very costly (in a wide sense) or even impossible.\\

To face this situation, we propose to use agent-based simulation models, which can provide a wealth of data to experiment and to test. 
Computer simulations are not free of controversy \citep[see][and references therein]{EdmMey13a, Saa17a} and one can rightfully argue that a 
mobile telecommunication network and the behaviours of its subscribers and the human population, in general, is too complex to be modelled in an agent-based simulation. This is basically the distinction between \emph{verification} and \emph{validation} \citep{Win09a} and it does not invalidate the utility of computer simulations.\\ 

A strong point for simulations is the provision of a synthetic ground truth \emph{for each parameterization}, thus allowing the analyst to 
investigate and test robust methods valid for a wide range of potential situations (never known in real conditions). A necessary element is 
to provide scenarios which are realistic enough, which can be undertaken by incorporating real metadata into the simulations (e.g. realistic network configurations and mobility patterns). This points once more at a necessary collaboration with data holders not in terms of data transfers or data transmissions at this stage, but to share critical information for the creating of an industrialised statistical process for the public good.

The use of simulation models under these conditions, in particular, with mobile network data allows us also to tackle methodological questions in a (semi-)empirical way which otherwise would consume many resources. Take for example the recent study by \citet{Ogu21} empirically showing the fallacy of the closest antenna to estimate the geolocation of mobile devices, invalidating thus many proposals based on Voronoi tessellations of the geographical territory of analysis (with implications also for the disclosure of information and the reidentification of individuals). Shortcomings from this simple use of Voronoi tessellations was already identified as early as 2015 by \citet{RicWidCraPan15a}. This kind of situation can be replicated and thus analysed with detail using the simulator presented here.\\

Therefore, for the development of statistical methodology, we see immediate advantages of using simulation models: 
(i) to have actual ground truth figures allowing us to conduct a thorough performance assessment of methods and parameters and a better 
understanding by comparison between actual population counts and their estimates, (ii) to identify different concrete aspects of the problem by configuring different scenarios in order to propose specific elements in the methodology to deal with them, (iii) to avoid the issue about the access to real data (see above) and its consequences (lack of data, confidentiality and privacy risks, legal concerns,...), and (iv) to provide a body of technical knowledge to reach informed partnership agreements between statistical offices and data holders. Real data cannot provide these conditions for research.\\

In general terms, the use of agent-based simulation models enjoys attractive features for the research on novel statistical methodologies, 
which we should be aware about and which are shared in other scientific disciplines.

These advantages are complemented with the use of R to deploy in a fast and trustworthy manner those software tools needed to test the novel methods. R can deal with many different features regarding these new data sources, namely, the geospatial representation of information, the visualization of combined information, the use of different statistical methods (statistical learning models, state space models, \dots) and reporting.\\

In this line, we have developed the \textsf{simutils} package to manage all the geospatial information around these specific data simulations. Our package provides means to parameterize and run simulations using and external tool specially designed for this purpose, to manage the output data of the simulation building geospatial objects needed to carry on specific analyzes and to compute some aggregates. We showed concrete examples of the main functionalities using data sets included in the package. 


\begin{thebibliography}{65}
	\providecommand{\natexlab}[1]{#1}
	\providecommand{\url}[1]{\texttt{#1}}
	\expandafter\ifx\csname urlstyle\endcsname\relax
	\providecommand{\doi}[1]{doi: #1}\else
	\providecommand{\doi}{doi: \begingroup \urlstyle{rm}\Url}\fi
	
	\bibitem[Ahas et~al.(2007)Ahas, Aasa, Silm, and Tiru]{AhaAasSilTir07a}
	R.~Ahas, A.~Aasa, S.~Silm, and M.~Tiru.
	\newblock {Mobile Positioning Data in Tourism Studies and Monitoring: Case
		Study in Tartu, Estonia}.
	\newblock In M.~Sigala, L.~Mich, and J.~Murphy, editors, \emph{Information and
		Communication Technologies in Tourism 2007}, pages 119--128, Vienna, 2007.
	Springer.
	
	\bibitem[Aust(2021)]{pkg:depgraph}
	F.~Aust.
	\newblock \emph{depgraph: Plot package dependency graph}, 2021.
	\newblock URL \url{https://github.com/crsh/depgraph}.
	\newblock R package version 0.1.0.
	
	\bibitem[Baldacci et~al.(2021)Baldacci, Ricciato, and Wirthmann]{BalRicWir21a}
	E.~Baldacci, F.~Ricciato, and A.~Wirthmann.
	\newblock A {R}eflection on {T}he {R}e({U}se) of {N}ew {D}ata {S}ources for
	{O}fficial {S}tatistics.
	\newblock \emph{{R}evista \'{I}ndice}, 83:\penalty0 8--11, 2021.
	
	\bibitem[Baldwin and Clark(2000)]{BalCla00a}
	C.~Baldwin and K.~Clark.
	\newblock \emph{Design Rules (vol.1): The power of modularity}.
	\newblock MIT Press, 2000.
	
	\bibitem[Box(1979)]{Box79a}
	G.~Box.
	\newblock Some problems of statistics and everyday life.
	\newblock \emph{Journal of the American Statistical Association}, 74:\penalty0
	1--4, 1979.
	
	\bibitem[Chambers(2008)]{Cha08a}
	J.~Chambers.
	\newblock \emph{Software for data analysis}.
	\newblock Springer, 2008.
	
	\bibitem[Chambers(2009)]{Cha09c}
	J.~Chambers.
	\newblock {F}acets of {R}.
	\newblock \emph{The R Journal}, 1\penalty0 (1):\penalty0 5--8, 2009.
	
	\bibitem[Davis(1995)]{Dav95}
	A.~M. Davis.
	\newblock Software prototyping.
	\newblock In \emph{Advances in computers}, volume~40, pages 39--63. Elsevier,
	1995.
	
	\bibitem[DeAngelis(2018)]{DeAng18}
	D.~L. DeAngelis.
	\newblock \emph{Individual-based models and approaches in ecology: populations,
		communities and ecosystems}.
	\newblock CRC Press, 2018.
	
	\bibitem[DGINS(2013)]{SchMem13a}
	DGINS.
	\newblock The {S}cheveningen {M}emorandum.
	\newblock Technical report, European Union, 2013.
	\newblock
	https://ec.europa.eu/eurostat/cros/news/scheveningen-memorandum-big-data-and-official-statistics-adopted-essc\_en.
	
	\bibitem[DGINS(2018)]{BucMem18a}
	DGINS.
	\newblock The {B}ucharest {M}emorandum.
	\newblock Technical report, European Union, 2018.
	\newblock
	https://ec.europa.eu/eurostat/web/european-statistical-system/-/dgins2018-bucharest-memorandum-adopted.
	
	\bibitem[Dowle and Srinivasan(2021)]{pkg:data.table}
	M.~Dowle and A.~Srinivasan.
	\newblock \emph{data.table: Extension of `data.frame`}, 2021.
	\newblock URL \url{https://CRAN.R-project.org/package=data.table}.
	\newblock R package version 1.14.2.
	
	\bibitem[Edmonds and Meyer(2013)]{EdmMey13a}
	B.~Edmonds and R.~Meyer, editors.
	\newblock \emph{{Simulating Social Complexity: A Handbook}}.
	\newblock Springer, 2013.
	
	\bibitem[{ESSnet~on~Big~Data}(2021)]{ESSnetBDMobile21a}
	{ESSnet~on~Big~Data}.
	\newblock {Software for Mobile Network Data in Official Statistics}, 2021.
	\newblock https://github.com/MobilePhoneESSnetBigData/.
	
	\bibitem[Esteban et~al.(2018)Esteban, Nov\'{a}s, {n}a, Salgado, and
	Sanguiao]{Esteban2018}
	E.~Esteban, M.~Nov\'{a}s, S.~S. {n}a, D.~Salgado, and L.~Sanguiao.
	\newblock Data organisation and process design based on functional modularity
	for a standard production process.
	\newblock \emph{Journal of Official Statistics}, 34\penalty0 (4):\penalty0
	811--833, 2018.
	
	\bibitem[{E}urostat(2014)]{ESSVIP}
	{E}urostat.
	\newblock {V}ision 2020 {I}mplementation {P}ortfolio, 2014.
	\newblock URL
	\url{http://ec.europa.eu/eurostat/web/ess/about-us/ess-vision-2020/implementation-portfolio}.
	
	\bibitem[Eurostat(2021)]{ESSnetBD21a}
	Eurostat.
	\newblock {ESSnet on Big Data}, 2021.
	\newblock https://ec.europa.eu/eurostat/cros/content/essnet-big-data-1\_en.
	
	\bibitem[{E}urostat(2021)]{EurBD21a}
	{E}urostat.
	\newblock {B}ig {D}ata, 2021.
	\newblock URL \url{https://ec.europa.eu/eurostat/cros/content/big-data\_en}.
	\newblock Collaboration in Research and Methodology for Official Statistics.
	
	\bibitem[Field(2015)]{Fie15a}
	E.~Field.
	\newblock {\textquotedblleft All Models Are Wrong, but Some Are
		Useful\textquotedblright}.
	\newblock \emph{Seismological Research Letters}, 86\penalty0 (2A):\penalty0
	291--293, 2015.
	
	\bibitem[Gelman et~al.(2013)Gelman, Carlin, Stern, Dunson, Vehtari, and
	Rubin]{GelCarSteDunVehRub13a}
	A.~Gelman, B.~Carlin, H.~Stern, D.~Dunson, A.~Vehtari, and D.~Rubin.
	\newblock \emph{Bayesian data analysis}.
	\newblock CRC Press, 2013.
	
	\bibitem[Gilbert and Troitzsch(2005)]{Gil05}
	N.~Gilbert and K.~Troitzsch.
	\newblock \emph{Simulation for the social scientist}.
	\newblock McGraw-Hill Education (UK), 2005.
	
	\bibitem[Grimm et~al.(2006)Grimm, Berger, Bastiansen, Eliassen, Ginot, Giske,
	Goss-Custard, Grand, Heinz, Huse, et~al.]{Grimm06}
	V.~Grimm, U.~Berger, F.~Bastiansen, S.~Eliassen, V.~Ginot, J.~Giske,
	J.~Goss-Custard, T.~Grand, S.~K. Heinz, G.~Huse, et~al.
	\newblock A standard protocol for describing individual-based and agent-based
	models.
	\newblock \emph{Ecological modelling}, 198\penalty0 (1-2):\penalty0 115--126,
	2006.
	
	\bibitem[Hand(2018)]{Han18a}
	D.~Hand.
	\newblock Statistical challenges of administrative and transaction data.
	\newblock \emph{Journal of the Royal Statistical Society A}, 181:\penalty0
	555--605, 2018.
	
	\bibitem[Hastie et~al.(2009)Hastie, Tibshirani, and Friedman]{HasTbiFri09a}
	T.~Hastie, R.~Tibshirani, and J.~Friedman.
	\newblock \emph{The Elements of Statistical Learning}.
	\newblock Springer, New York, 2009.
	
	\bibitem[HLG-MOS(2011)]{HLG11a}
	HLG-MOS.
	\newblock {S}trategic vision of the {H}igh-{L}evel {G}roup for strategic
	developments in business architecture in {S}tatistics.
	\newblock \emph{Conference of European Statisticians, Geneva, 14-16 June},
	2011.
	
	\bibitem[Kitchin(2015)]{Kit15a}
	R.~Kitchin.
	\newblock Big {D}ata and {O}fficial {S}tatistics: {O}pportunities, challenges
	and risks.
	\newblock \emph{Statistical Journal of the IAOS}, 31:\penalty0 471--481, 2015.
	
	\bibitem[Kitchin(2017)]{Kit17a}
	R.~Kitchin.
	\newblock Thinking critically about and researching algorithms.
	\newblock \emph{Information, Communication \& Society}, 20:\penalty0 14--29,
	2017.
	
	\bibitem[Law(2008)]{Law08a}
	A.~Law.
	\newblock {How To Build Valid and Credible Simulation Models}.
	\newblock In S.~Mason, R.~Hill, L.M\"{o}nch, O.~Rose, T.~Jefferson, and
	J.~Fowler, editors, \emph{Proceedings of the 2008 Winter Simulation
		Conference}, pages 39--47, 2008.
	
	\bibitem[Lovelace et~al.(2022)Lovelace, Nowosad, and Muenchow]{LovNowMue22a}
	R.~Lovelace, J.~Nowosad, and J.~Muenchow.
	\newblock \emph{{Geocomputation with R}}.
	\newblock CRC Press, 2022.
	\newblock https://geocompr.robinlovelace.net/.
	
	\bibitem[Miao et~al.(2016)Miao, Zander, Sung, and Slimane]{MiaZanSunSli16a}
	G.~Miao, J.~Zander, K.~Sung, and S.~Slimane.
	\newblock \emph{{Fundamental of Mobile Data Networks}}.
	\newblock Cambridge University Press, 2016.
	
	\bibitem[Munoz et~al.(2009)Munoz, Bouchereau, Vargas, and
	Enr\'{\i}quez-Caldera]{MunBouVarEnrCal09}
	D.~Munoz, F.~Bouchereau, C.~Vargas, and R.~Enr\'{\i}quez-Caldera.
	\newblock \emph{{Position Location Techniques and Applications}}.
	\newblock Academic Press, Burlington, 2009.
	
	\bibitem[Murphy(2013)]{Mur13a}
	K.~Murphy.
	\newblock \emph{Machine learning: a probabilistic perspective}.
	\newblock MIT Press, 2013.
	
	\bibitem[Oancea(2021)]{Simulator21a}
	B.~Oancea.
	\newblock {Mobile Network Data Simulator}, 2021.
	\newblock https://github.com/bogdanoancea/simulator.
	
	\bibitem[Oancea et~al.(2019)Oancea, Necula, Salgado, Sanguiao, and
	Barragán]{Oan19}
	B.~Oancea, M.~Necula, D.~Salgado, L.~Sanguiao, and S.~Barragán.
	\newblock Deliverable i.2 (data simulator) a simulator for network event data.
	\newblock Technical report, Eurostat, 2019.
	\newblock URL
	\url{https://ec.europa.eu/eurostat/cros/sites/default/files/WPI_Deliverable_I2_Data_Simulator_-_A_simulator_for_network_event_data.pdf}.
	
	\bibitem[Oancea et~al.(2020)Oancea, Barragán, Sanguiao, and Salgado]{Oan20}
	B.~Oancea, S.~Barragán, L.~Sanguiao, and D.~Salgado.
	\newblock Deliverable i.4 (information techonologies) some it tools for the
	production of official statistics with mobile network data.
	\newblock Technical report, Eurostat, 2020.
	\newblock URL
	\url{https://ec.europa.eu/eurostat/cros/system/files/wpi_deliverable_i3_a_proposed_production_framework_with_mobile_network_data_2020_11_26_final.pdf}.
	
	\bibitem[Oancea et~al.(2021)Oancea, Salgado, Sanguiao, and
	Barragán]{oancearsr2021}
	B.~Oancea, D.~Salgado, L.~Sanguiao, and S.~Barragán.
	\newblock A {Se}t of {R} {P}ackages to {E}stimate {P}opulation {C}ounts from
	{M}obile {P}hone {D}ata.
	\newblock \emph{{R}omanian {S}tatistical {R}eview}, \penalty0 (1):\penalty0
	17--38, 2021.
	
	\bibitem[Ogulenko et~al.(2021{\natexlab{a}})Ogulenko, Benenson, Omer, and
	Alon]{OguBenOmeAlo21a}
	A.~Ogulenko, I.~Benenson, I.~Omer, and B.~Alon.
	\newblock Probabilistic positioning in mobile phone network and its
	consequences for the privacy of mobility data.
	\newblock \emph{Computers, Environment and Urban Systems}, 85:\penalty0 101550,
	2021{\natexlab{a}}.
	\newblock ISSN 0198-9715.
	\newblock \doi{https://doi.org/10.1016/j.compenvurbsys.2020.101550}.
	\newblock URL
	\url{https://www.sciencedirect.com/science/article/pii/S0198971520302830}.
	
	\bibitem[Ogulenko et~al.(2021{\natexlab{b}})Ogulenko, Benenson, Toger,
	{\"O}sth, and Siretskiy]{Ogu21}
	A.~Ogulenko, I.~Benenson, M.~Toger, J.~{\"O}sth, and A.~Siretskiy.
	\newblock The fallacy of the closest antenna: Towards an adequate view of
	device location in the mobile network.
	\newblock \emph{arXiv preprint arXiv:2109.02154}, 2021{\natexlab{b}}.
	
	\bibitem[O'Sullivan and Perry(2013)]{OSul13}
	D.~O'Sullivan and G.~L. Perry.
	\newblock \emph{Spatial simulation: exploring pattern and process}.
	\newblock John Wiley \& Sons, 2013.
	
	\bibitem[Pebesma(2018)]{pkg:sf}
	E.~Pebesma.
	\newblock {Simple Features for R: Standardized Support for Spatial Vector
		Data}.
	\newblock \emph{{The R Journal}}, 10\penalty0 (1):\penalty0 439--446, 2018.
	\newblock \doi{10.32614/RJ-2018-009}.
	\newblock URL \url{https://doi.org/10.32614/RJ-2018-009}.
	
	\bibitem[Pebesma(2021)]{pkg:stars}
	E.~Pebesma.
	\newblock \emph{stars: Spatiotemporal Arrays, Raster and Vector Data Cubes},
	2021.
	\newblock URL \url{https://CRAN.R-project.org/package=stars}.
	\newblock R package version 0.5-3.
	
	\bibitem[Ricciato et~al.(2015)Ricciato, Widhalm, Craglia, and
	Pantisano]{RicWidCraPan15a}
	F.~Ricciato, P.~Widhalm, M.~Craglia, and F.~Pantisano.
	\newblock {Estimating population density distribution from network-based mobile
		phone data}.
	\newblock Technical report, Joint Research Centre, 2015.
	\newblock Technical Report JRC96568. Available at
	https://ec.europa.eu/jrc/file/document/26733.
	
	\bibitem[Robert(2016)]{pkg:DependenciesGraphs}
	T.~Robert.
	\newblock \emph{DependenciesGraphs: Dependencies visualization between
		functions and environments}, 2016.
	\newblock URL \url{https://github.com/datastorm-open/DependenciesGraphs}.
	\newblock R package version 0.3.
	
	\bibitem[Saam(2017)]{Saa17a}
	N.~Saam.
	\newblock {What is a Computer Simulation? A Review of a Passionate Debate}.
	\newblock \emph{{Journal of General Philosophy of Science}}, 48:\penalty0
	293–--309, 2017.
	\newblock \doi{10.1007/s10838-016-9354}.
	\newblock URL \url{https://doi.org/10.1007/s10838-016-9354}.
	
	\bibitem[Salgado and Oancea(2020)]{SalOan20a}
	D.~Salgado and B.~Oancea.
	\newblock On new data sources for the production of official statistics, 2020.
	\newblock arXiv:2003.06797v1.
	
	\bibitem[Salgado et~al.(2021)Salgado, Sanguiao, Oancea, Barrag\'{a}n, and
	Necula]{SalSanOanBarNec21a}
	D.~Salgado, L.~Sanguiao, B.~Oancea, S.~Barrag\'{a}n, and M.~Necula.
	\newblock {An end-to-end statistical process with mobile network data for
		official statistics}.
	\newblock \emph{{EPJ Data Science}}, 10:\penalty0 20, 2021.
	\newblock \doi{10.1140/epjds/s13688-021-00275-w}.
	\newblock URL
	\url{https://epjdatascience.springeropen.com/articles/10.1140/epjds/s13688-021-00275-w}.
	
	\bibitem[Sanguiao et~al.(2020)Sanguiao, Salgado, and Oancea]{pkg:destim}
	L.~Sanguiao, D.~Salgado, and B.~Oancea.
	\newblock \emph{destim: R package for mobile devices position estimation},
	2020.
	\newblock URL \url{https://github.com/Luis-Sanguiao/destim}.
	\newblock R package version 0.1.0.
	
	\bibitem[S\"{a}rndal et~al.(1992)S\"{a}rndal, Swensson, and
	Wretman]{SarSweWre92a}
	C.-E. S\"{a}rndal, B.~Swensson, and J.~Wretman.
	\newblock \emph{Model assisted survey sampling}.
	\newblock Springer, New York, 1992.
	
	\bibitem[Schabenberger and Gotway(2005)]{SchGot05a}
	O.~Schabenberger and C.~Gotway.
	\newblock \emph{Statistical {M}ethods for {S}patial {D}ata {A}nalysis}.
	\newblock Texts in Statistical Science. Chapman \& Hall/CRC, Boca Raton, 2005.
	
	\bibitem[ten Bosch et~al.(2020)ten Bosch, van~der Loo, and
	Kowarik]{BosLooKow20a}
	O.~ten Bosch, M.~van~der Loo, and A.~Kowarik.
	\newblock The awesome list of official statistics software: 100 \dots and
	counting.
	\newblock The Use of R in Official Statistics Conference, 2-4 December 2020.
	See also https://github.com/SNStatComp/awesome-official-statistics-software,
	2020.
	
	\bibitem[Tennekes and Gootzen(2021)]{TenGoo21a}
	M.~Tennekes and Y.~Gootzen.
	\newblock {A Bayesian approach to location estimation of mobile devices from
		mobile network operator data}, 2021.
	\newblock arXiv:2110.00439v1.
	
	\bibitem[Tennekes et~al.(2020)Tennekes, Gootzen, and Shah]{TenGooSha20a}
	M.~Tennekes, Y.~Gootzen, and S.~Shah.
	\newblock {A Bayesian approach to location estimation of mobile devices from
		mobile network operator data}.
	\newblock Technical report, CBS/Statistics Netherlands, 2020.
	\newblock Working Paper 06-20. Available at
	https://www.cbs.nl/-/media/\_pdf/2020/22/cbds\_working\_paper\_location\_estimation.pdf.
	
	\bibitem[Tesfatsion(2002)]{Tes02}
	L.~Tesfatsion.
	\newblock Agent-based computational economics: Growing economies from the
	bottom up.
	\newblock \emph{Artificial life}, 8\penalty0 (1):\penalty0 55--82, 2002.
	
	\bibitem[Ucar et~al.(2021)Ucar, Gramaglia, Fiore, Smoreda, and
	Moro]{UcaGraFioSmoMor21a}
	I.~Ucar, M.~Gramaglia, M.~Fiore, Z.~Smoreda, and E.~Moro.
	\newblock {News or social media? Socio-economic divide of mobile service
		consumption}.
	\newblock \emph{Journal of The Royal Society Interface}, 18:\penalty0 20210350,
	2021.
	\newblock \doi{10.1098/rsif.2021.0350}.
	\newblock URL
	\url{https://royalsocietypublishing.org/doi/10.1098/rsif.2021.0350}.
	
	\bibitem[UNECE(2019{\natexlab{a}})]{GAMSO12}
	UNECE.
	\newblock Generic activity model for statistical organizations v1.2,
	2019{\natexlab{a}}.
	\newblock https://statswiki.unece.org/display/GAMSO/GAMSO+v1.2.
	
	\bibitem[UNECE(2019{\natexlab{b}})]{GSBPM51}
	UNECE.
	\newblock Generic {S}tatistical {B}usiness {P}rocess {M}odel v5.1,
	2019{\natexlab{b}}.
	\newblock https://statswiki.unece.org/display/GSBPM/GSBPM+v5.1.
	
	\bibitem[UNECE(2019{\natexlab{c}})]{GSIM12}
	UNECE.
	\newblock Generic statistical information model v1.2, 2019{\natexlab{c}}.
	\newblock https://statswiki.unece.org/display/gsim.
	
	\bibitem[UNECE(2021)]{HLG21}
	UNECE.
	\newblock High-level group for the modernisation of statistical production and
	services, 2021.
	\newblock
	https://unece.org/statistics/networks-of-experts/high-level-group-modernisation-statistical-production-and-services.
	
	\bibitem[{UNECE}(2021)]{UNBD21a}
	{UNECE}.
	\newblock {B}ig {D}ata, 2021.
	\newblock URL \url{https://unece.org/statistics/ces/big-data}.
	
	\bibitem[Urbanek(2021)]{pkg:rJava}
	S.~Urbanek.
	\newblock \emph{rJava: Low-Level R to Java Interface}, 2021.
	\newblock URL \url{https://CRAN.R-project.org/package=rJava}.
	\newblock R package version 1.0-5.
	
	\bibitem[Valliant et~al.(2000)Valliant, Dorfmann, and Royall]{ValDorRoy00a}
	R.~Valliant, A.~Dorfmann, and R.~Royall.
	\newblock \emph{Finite population sampling and inference. A prediction
		approach}.
	\newblock Wiley, New York, 2000.
	
	\bibitem[van~der Loo(2019)]{Loo19}
	M.~P. van~der Loo.
	\newblock \emph{Systematic approaches to data validation and data cleaning for
		statistical production}, 2019.
	\newblock Seminar at Statistics Spain (INE).
	
	\bibitem[van~der Loo(2021)]{Loo21}
	M.~P. van~der Loo.
	\newblock {A Method for Deriving Information from Running R Code}.
	\newblock \emph{{The R Journal}}, 13\penalty0 (1):\penalty0 42--52, 2021.
	\newblock \doi{10.32614/RJ-2021-056}.
	\newblock URL \url{https://doi.org/10.32614/RJ-2021-056}.
	
	\bibitem[Winsberg(2009)]{Win09a}
	E.~Winsberg.
	\newblock {Computer Simulation and the Philosophy of Science}.
	\newblock \emph{{Philosophy Compass}}, 4/5:\penalty0 835–--845, 2009.
	\newblock \doi{10.1111/j.1747-9991.2009.00236.x}.
	\newblock URL \url{https://doi.org/10.1111/j.1747-9991.2009.00236.x}.
	
	\bibitem[Wit et~al.(2012)Wit, van~den Heuvel, and Romeijn]{WitHeuRom12a}
	E.~Wit, E.~van~den Heuvel, and J.-W. Romeijn.
	\newblock {‘All models are wrong\dots’: an introduction to model
		uncertainty}.
	\newblock \emph{Statistica Neerlandica}, 66\penalty0 (3):\penalty0 217--236,
	2012.
	
\end{thebibliography}

\end{document}